# Blissful (A)Ignorance:

People form overly positive impressions of others based on their
written messages, despite wide-scale adoption of Generative AI

Jiaqi Zhu[a] & Andras Molnar[b*]

[a]University of Michigan, Department of Statistics
1085 South University Ann Arbor, MI 48109-1107
Email: jiaqizhu@umich.edu

[b]University of Michigan, Department of Psychology
530 Church Street Ann Arbor, MI 48109–1043
Email: andrasm@umich.edu
[*]corresponding author

## Declaration of interests

The authors declare that they have no known competing financial interests or personal
relationships that could have appeared to influence the work reported in this paper.

# Blissful (A)Ignorance:

People form overly positive impressions of others based on their written messages,

despite wide-scale adoption of Generative AI

## Abstract

As the use of Generative AI (GenAI) tools becomes more prevalent in interpersonal communication, understanding their impact on social perceptions is crucial. According to signaling theory, GenAI may undermine the credibility of social signals conveyed in writing, since it reduces the cost of writing and makes it hard to verify the authenticity of messages. Using a pre-registered large-scale online experiment ($N = 647$; Prolific), featuring scenarios in a range of communication contexts (personal vs. professional; close others vs. strangers), we explored how senders' use of GenAI influenced recipients' impressions of senders, both when GenAI use was known or uncertain. Consistent with past work, we found strong negative effects on social impressions when disclosing that a message was AI-generated, compared to when the same message was human-written. However, under the more realistic condition when potential GenAI use was not explicitly highlighted, recipients did not exhibit *any* skepticism towards senders, and these "uninformed" impressions were virtually indistinguishable from those of fully human-written messages. Even when we highlighted the potential (but uncertain) use of GenAI, recipients formed overly positive impressions. These results are especially striking given that 46% of our sample admitted having used such tools for writing messages, just within the past two weeks. Our findings put past work in a new light: While social judgments can be substantially affected when GenAI use is explicitly disclosed, this information may not be readily available in more realistic communication settings, making recipients blissfully ignorant about others' potential use of GenAI.

## Highlights

- Examine the impact of GenAI use on social impressions in written communication
- Large-scale online experiment (N= 647; Prolific) featuring 4 hypothetical scenarios
- People form negative impressions when they know that messages are AI-generated
- When uncertain or uninformed about AI use, people form overly positive impressions
- People remain blissfully ignorant, despite 46% having used GenAI to write messages

# 1. Introduction

According to a nationally representative survey conducted in August 2024, about 39% of the U.S. population aged 18-64 is already using Generative AI, most commonly ChatGPT, with 10.6% reporting daily usage at work (Bick et al., 2024). As Bick et al. (2024) highlight, the adoption of Generative AI has been quicker than that of personal computers or the Internet. This rapid adoption has been widely documented across age groups, disciplines, and occupations, including but not limited to office workers (Humlum & Vestergaard, 2024), middle and high school students (Zhu et al., 2024), STEM researchers (Van Noorden & Perkel, 2023), humanities researchers (Dedema & Ma, 2024), medical students (Zhang et al., 2024), and medical practitioners (Blease et al., 2024).

In addition to the widespread adoption of Generative AI for personal and professional use, these technologies also have the potential to transform *interpersonal interactions*. Generative AI tools empower users with the ability to generate high-quality, complex, personalized, and contextually relevant written content with minimal effort. They have the capacity to enhance the efficiency and quality of social interactions, for example, by boosting the use of positive emotional language (Hohenstein et al., 2023).

At the same time, suspected AI-use may have adverse effects on social relationships (e.g., Glikson & Asscher, 2023; Hohenstein et al., 2023; Lim et al., 2025; Weiss et al., 2022), and instead of building social connections and facilitating interactions, these emerging technologies may erect new barriers to human-to-human cooperation and coordination. Wojtowicz and DeDeo (2025) discuss these negative effects by highlighting how Generative AI may undermine "mental proofs." Mental proofs are observable actions that allow audiences to make inferences about the communicator's hidden mental states such as their intentions, goals, and values. As Wojtowich and DeDeo (2025) argue, outsourcing the laborious process of writing to an algorithm—thereby making social interactions more efficient and less effortful—may lead to the paradoxical consequence of undermining trust and coordination between people, because it weakens the link between what people communicate and what people actually think, feel, or want.

But are people actually becoming more skeptical towards others when judging them based on their written messages? Do people respond to the increasing prevalence of Generative AI use by evaluating written content more critically, or do they remain blissfully ignorant to the possibility that messages may be AI-generated? How does (the lack of) information about AI use affect social impressions, especially in more realistic settings when there is a fundamental uncertainty about others' use of AI? In the present paper, we seek answers to these questions and experimentally investigate the current "state of skepticism" in interpersonal communication.

## 1.1. Signaling theory: writing as a source of credible social signals

Written communication is a rich source of social signals, as every choice a writer makes—words, tone, and sentence structure—reflects the underlying characteristics of the individual (e.g., Mairesse et al., 2007; Pennebaker et al., 2003). These signals enable readers to assess the personal traits of the writer, such as trustworthiness, thoughtfulness, diligence, and kindness, or conversely, deceitfulness, superficiality, laziness, and antagonism. Such inferences allow people to decide whether to cooperate, compete, approach, or avoid others. For example, cover letters help employers choose which candidates to interview; dating profiles let singles decide whether they want to meet a potential date; apology letters may determine whether someone is forgiving or holding onto a grudge; and academic publications may serve as the primary avenue for researchers to showcase their intellect, creativity, and effort.

However, signals can be faked, especially when communicators have a strategic interest in deceiving or manipulating their audience. A job candidate may ask a seasoned colleague to write a cover letter on their behalf; an apology letter may be based on a template; and an academic publication may plagiarize past work. According to signaling theory (see, e.g., Connelly et al., 2011; Spence 1973, 2002), the usefulness of signals—any signal, not just in written communication—critically depends on the credibility of signaling, that is, how likely a signal is honest or fake. A core prediction of signaling theory is that as the *cost* of signaling increases—which can be time, effort, emotional cost, or even reputation—it becomes more credible (Spence 1973, 2002). Therefore, observers are constantly monitoring costs to gauge the credibility of signals (Gintis et al., 2001; also see "strategic vigilance", Heintz et al., 2016). As Chaudhry & Wald (2022) highlight, three qualities can make an observable signal to be perceived as costly: 1) it is difficult-to-fake; 2) verifiable; and 3) self-sacrificing. For example, a handwritten note of apology is a more credible signal of true remorse than a text message, as hand-writing is more difficult-to-fake, easier to verify, and takes more effort and time to produce (i.e., requires more "self-sacrifice") than texting someone.

## 1.2. How Generative AI may undermine the social signaling function of written messages

Generative AI may undermine all three of these qualities of written signals, substantially reducing their perceived cost and credibility. First, these tools offer users unprecedented flexibility in generating any type of written content, editing style and grammar, and tailoring messages to all sorts of audiences, thereby making messages extremely easy to fake. To make matters worse, people are largely incapable of telling the difference between AI-generated and human-created content (e.g., Jakesch et al., 2023; Köbis & Mossink, 2021; Kreps et al., 2022; Porter & Machery, 2024). Second, AI-generated content is difficult, if not impossible, to verify. Unlike more "traditional" forms of fake signals in written communication—such as plagiarism or ghostwriting—which can be effectively identified as fake, we generally lack the tools (either algorithmic or heuristic) to reliably determine if, and to what extent, Generative AI was involved in writing a message (Sadasivan et al., 2023; Tang et al., 2023). Detecting AI-generated content is especially challenging when the text is a mix of AI-generated and human-written content ("mixtext", see Gao et al., 2024). Finally, these technologies have significantly reduced the cost of writing messages—not only in terms of direct costs incurred by the communicator (time and effort), but also in terms of overall *energy consumption.* According to Tomlinson et al. (2024)'s estimates, tools like ChatGPT have made writing so effortless and efficient that these systems emit 130-1500 times less $CO_2$/page of text than human writers.

There is already some empirical evidence that documents the adverse effect of Generative AI on social impressions and interactions. For example, Glikson & Asscher (2023) investigated the effectiveness of apologies, depending on whether the communicator used AI while writing an apology message. Participants rated the apology written with the help of AI as less authentic, less sincere, and as a result, were less likely to forgive the communicator. These negative effects were mitigated by the limited use of AI (i.e., when used for translation and/or correcting grammatical mistakes only). Lim et al. (2025) found similar negative effects on the perceived sincerity of corporate apologies. In their experiment, participants read about a fictitious company issuing an apology for a data breach incident. The authors found that when the company's press release was "generated with the assistance of ChatGPT", participants rated these statements as less sincere and less empathetic, and consequently, reported lower willingness to forgive the incident. Other related work looked at social impressions more broadly. For example, Weiss et al. (2022) found that the disclosed use of AI in job applicants' messages negatively affected their interpersonal perceptions. Hohenstein et al. (2023) examined how incorporating "smart replies" in messaging affected social perceptions and found that people evaluate their conversational partners more negatively when they think that their counterpart used algorithmic responses.

## 1.3. The role of uncertainty and attention in social judgments

There are, however, at least two crucial differences between the stylized interactions studied in this past line of research and real situations. First, people in real situations almost never know with certainty whether (and to what extent) a communicator has used Generative AI. Assuming that communicators are aware of the potential negative consequences of disclosing their use of AI, they are *motivated to conceal* all AI involvement. This rational strategy of concealing AI use, combined with the difficulty of detecting and distinguishing AI-generated content (Jakesch et al., 2023; Köbis & Mossink, 2021; Kreps et al., 2022), forces audiences to rely on probabilistic inferences instead. That is, in most situations, people must form impressions of others and make decisions under the fundamental uncertainty of not knowing whether the communicator has used Generative AI, let alone, to what extent.

One goal of this paper is to study such judgments under uncertainty: when people are *aware* of the possibility that a message might have been generated by AI but have no way of knowing this. According to a simple decision-theoretical model, social impressions under uncertainty, $I_U$, are the probability-weighted combination of impressions under certainty:

$$I_U = p_H \, I_H + p_{AI} \, I_{AI}$$

*Equation 1.* Social impression under uncertainty where $I_H$ is the social impression of someone who is *known* to have written a message entirely on their own, $I_{AI}$ is the impression of someone who is *known* to have generated a message entirely by using AI, while $p_H$ and $p_{AI}$ are the corresponding probabilities ($p_{AI} = 1 - p_H$).[1] Assuming that $I_H > I_{AI}$ (aligned with prior empirical evidence), the above model implies that social impressions would turn increasingly more negative (positive) as people would become more (less) suspicious of others, as a function of $p_{AI}$.

---

[1] We discuss the validity and plausibility of these assumptions in Section 4.2.

In addition to uncertainty, there is another—in our view, even more fundamental—difference between real situations and the stylized interactions studied in past work: *selective attention* to decision-relevant information, specifically, attention to the possibility that a message may have been generated by AI. Decision-makers face attentional constraints (see "bounded rationality", Simon, 1955; Slovic, 1972), and as Kahneman (2011) highlighted, our understanding of a situation is often constrained by its presentation and what is most salient in our minds ("What you see is all there is"; also see Enke, 2020). Consequently, relevant factors may be overlooked during decision-making unless they are easily accessible and salient enough.

As a recent example for this phenomenon, Molnar et al. (2023) investigated how people judged potentially risky behaviors during the height of the COVID-19 pandemic in the U.S. (May 2021), depending on the information they knew about a target person's vaccination status. Since some behaviors (e.g., dining indoors, visiting a crowded bar) were deemed as riskier and more inappropriate if someone was engaging in these without being vaccinated against COVID-19, the authors expected that observers' judgments would be sensitive to information about vaccination status and people would consider the probability that someone else was vaccinated or not. Indeed, when people knew with certainty that someone was vaccinated or not, they judged the behavior of the former person as less risky and more appropriate. However, when people simply observed and evaluated someone engaging in the same behaviors, without thinking about their vaccination status, they completely ignored the possibility that this person may not have been vaccinated and rated such encounters as equally low risk as those cases when they *knew* that the target was vaccinated. However, as soon as participants were prompted to *consider* vaccination status (without getting any specific information), their judgments became much more critical and aligned with the probabilistic decision-theoretic model depicted in Equation 1. This example highlights that people may ignore decision-relevant information even in high-stakes situations, when that information is not top of their mind or not highlighted.

Translating this insight into the context of written communication, it may be the case that, by default, people do not even pay attention to the possibility that a message may have been generated by AI. While existing studies are informative about judgments and behavior under certainty and when people already pay attention to the possibility that a message was generated by AI, they did not test whether people would actually pay attention to such information without being prompted to do so. By contrast, in the present work we compare participants' social impressions under more naturalistic circumstances and uncertainty: when they are uncertain, or even unaware, that someone may have used Generative AI to send them a message.

## 1.4. Research design and hypotheses

We investigate whether social impressions are affected by the *possibility* that a message was generated by AI, both when such possibility is implicit (not highlighted to participants) and explicitly highlighted in the decision-context. We also compare impressions in these fundamentally uncertain situations to fully informed impressions, that is, when participants know whether a message was written by a person or generated by AI. This approach not only allows us to investigate how the presence of uncertainty affects impressions but also to examine whether people have become more critical towards written content, as a natural response to the proliferation of Generative AI tools in human-to-human communication. Our results can shed light on whether people have already developed such natural skepticism or remain blissfully ignorant and largely fail to consider the possibility that written messages may be AI-generated.

To investigate these, we designed four hypothetical scenarios to cover a range of possible situations (e.g., in personal vs. workplace settings; between close others vs. strangers), in which participants received an email from someone else (see Section 2.3 and Appendix for details). Across four experimental conditions, we then manipulated the type of information participants received about the origin of this email. In the two certain conditions, we told participants that this message was either fully generated by AI or fully written by the sender. In the two other conditions, we either provided no information about the possibility of AI involvement or told participants that they are uncertain about the origin of the message (i.e., highlighted the possibility of AI involvement but didn't provide any specific information). We then measured participants' social impressions of the sender.

Based on past work (e.g., Glikson & Asscher, 2023; Hohenstein et al., 2023; Lim et al., 2025; Weiss et al., 2022), we first hypothesized that fully informed participants would judge the sender of the message more negatively if they knew that the sender used AI:

**H1. People form more negative social impressions of someone known to have used Generative AI in writing a message, compared to someone known to have written the same message entirely on their own, without the help of Generative AI.**

In addition, our experiment allowed us to test the following five research questions:[2]

> ***RQ2A. Do people form different social impressions of someone known to have used Generative AI in writing a message, compared to someone whose potential use of Generative AI is not highlighted?***

> ***RQ2B. Do people form different social impressions of someone known to have used Generative AI in writing a message, compared to someone whose potential use of Generative AI is highlighted but remains uncertain?***

> ***RQ3A. Do people form different social impressions of someone known to have written the message entirely on their own, compared to someone whose potential use of Generative AI is not highlighted?***

> ***RQ3B. Do people form different social impressions of someone known to have written the message entirely on their own, compared to someone whose potential use of Generative AI is highlighted but remains uncertain?***

> ***RQ4. Do people form different social impressions of someone whose potential use of Generative AI is not highlighted, compared to someone else whose potential use of Generative AI is highlighted but remains uncertain?***

Research questions RQ2A and RQ3A examine how the impressions formed under the most naturalistic conditions (i.e., simply evaluating someone based on their message, without being actively reminded or prompted to think about Generative AI) may differ from the impressions made by fully informed participants. Research questions RQ2B and RQ3B allow us to estimate participants' belief about the probability that a message was generated by AI, using the logic of the simple decision-theoretic model under uncertainty we described in Section 1.3 (see Equation 1). That is, a larger difference in RQ2B (and consequently, a smaller difference in RQ3B) would imply lower estimates for the probability that a message was generated by AI. Finally, RQ4 allows us to test the role of attention: whether people form different impressions if the possibility of AI use is highlighted, even when we don't provide any specific information about the particular target's AI use. Any significant difference between the two conditions included in RQ4 would imply that prompting people to think about Generative AI inherently changes how they form impressions, suggesting the role of selective attention to decision-relevant information.

---

[2] In the pre-registration, we referred to these research questions as "exploratory hypotheses" (see H2A-H4), but as indicated there, we did not make a priori directional predictions for these. Our analyses account for the exploratory nature of these research questions (i.e., we adjusted for multiple comparisons in null hypothesis testing).

## 2. Methods

### 2.1. Transparency and openness

We report how we determined our sample size (see Section 2.2), all data exclusions (if any), all manipulations, and all measures, and the studies follow JARS (Appelbaum et al., 2018). All data, analysis code, and research materials are publicly available at OSF: https://osf.io/xrpy9/?view_only=ba3a2e85211140fdb005ed21de92b2a7. We analyzed data using R, version 4.4.2 (R Core Team, 2024). We pre-registered the experiment on AsPredicted.org: https://aspredicted.org/38jb-s8dk.pdf. The study was reviewed and approved by the Institutional Review Board at [*name of institution and IRB reference number anonymized for peer review*]. All research was performed in accordance with relevant guidelines and regulations, and informed consent was obtained from all participants.

### 2.2. Sample

A priori power analysis using the G*Power 3.1.9.7 computer program (Faul et al., 2009) indicated that a sample of 164 participants per condition (pooled across scenarios) would be needed to detect a medium effect size ($d = 0.40$) with 95% power between any two conditions, using pairwise between-participants (two-tailed) $t$-tests. Based on this power analysis, we aimed to collect 160 responses per condition (i.e., 640 total), after data exclusions. Since each condition featured four different scenarios, we aimed to collect 40 responses in each scenario within each condition. We stopped data collection as soon as we reached these target sample sizes, after applying data exclusions. The data stopping rule was pre-registered.

We recruited 671 participants via Prolific (www.prolific.com). When recruiting participants, we set the following three eligibility criteria: 1) participants must be located in the United States; 2) speak English as their primary language; and 3) have an approval rate of 100 on Prolific. After we finished with recruitment, we excluded 24 participants (3.6%) from our analyses: 21 (3.1%) who quit the study before completing it and 3 (0.5%) who failed the attention check question. These exclusion criteria were also pre-registered. The final sample contained 647 participants (49.1% female; $M_{age}$ = 37.52 years). We report the detailed demographic information in Table 1.

*Table 1:* Summary of sample demographics

| Individual-level variables | *N* | Percent | Mean | SD |
|---|---|---|---|---|
| Age | 647 | | 37.52 | 12.18 |
| Gender | | | | |
|   Female | 318 | 49.1 | | |
|   Male | 312 | 48.2 | | |
|   Non-binary | 16 | 2.5 | | |
|   Prefer not to say | 1 | 0.2 | | |
| Highest level of education | | | | |
|   Some high school or less | 8 | 1.2 | | |
|   High school diploma or equivalent | 84 | 13.0 | | |
|   Some college or associate degree | 168 | 26.0 | | |
|   Bachelor's degree | 285 | 44.0 | | |
|   Master's degree | 77 | 11.9 | | |
|   Professional degree | 15 | 2.3 | | |
|   Doctoral degree | 9 | 1.4 | | |
|   Prefer not to say | 1 | 0.2 | | |
| Annual household income | | | | |
|   < $25,000 | 88 | 13.6 | | |
|   $25,000 - $49,999 | 126 | 19.5 | | |
|   $50,000 - $74,999 | 148 | 22.9 | | |
|   $75,000 - $99,999 | 82 | 12.7 | | |
|   $100,000 - $149,999 | 109 | 16.8 | | |
|   $150,000 or above | 79 | 12.2 | | |
|   Prefer not to say | 15 | 2.3 | | |

## 2.3. Procedure

We directed participants to a Qualtrics survey titled "Social Perceptions in Interpersonal Communication" and asked them to imagine a hypothetical scenario. Crucially, we did not tell participants in the instructions (or during recruitment) that this study would be about Generative AI to avoid prompting them to think about this technology. In the scenario, we presented participants with a brief background story and then told them that another person, Alex, had sent them an email. To capture a broad range of potential situations in which someone may be judged based on their email, we implemented a stimulus sampling method (see Wells & Windschitl, 1999). We randomly assigned participants to imagine one of the following four scenarios:

1) receiving a gratitude message from a close friend ("gratitude");
2) receiving an application from a candidate who is applying for a nanny position ("nanny");
3) receiving a cover letter from a job applicant for a data analyst position ("cover letter"); or
4) receiving feedback on a marketing project ("feedback").

We generated these messages using ChatGPT-4o and minimally edited them to have about the same length across contexts ($M = 232$ words; range: 217-243 words). We report the full text of these scenarios in the Appendix. Across the scenarios, we varied whether the sender, Alex, was someone who is socially close to the participant (i.e., their friend in the gratitude scenario or their colleague in the feedback scenario) or socially distant (i.e., a stranger in the nanny and cover letter scenarios), and whether the context was more personal (i.e., gratitude and nanny scenarios) as opposed to professional (i.e., cover letter and feedback scenarios).

After participants read the message sent by Alex, we implemented the main experimental manipulation: Across four conditions, we varied what participants knew about how Alex created the email. In the "no information" condition, that we designed to most closely resemble naturalistic communication settings, we didn't provide any further information, and participants simply proceeded to evaluate Alex based on their message.

In the other three conditions, we first asked participants to read a brief description of Generative AI chatbots—what these tools are and what they can be used for (see the Appendix)—then we informed participants whether Alex used a Generative AI chatbot to create their message.

In the "human" condition, we told participants that *"Alex has written their message without using a Generative AI Chatbot. Alex wrote the text on their own, word by word."*

In the "AI" condition, participants learned that *"Alex has generated their entire message by using a Generative AI Chatbot. Alex didn't change or modify the AI-generated text at all.*"

Finally, in the "uncertain" condition, we asked participants to imagine that they *"are uncertain: whether Alex has generated their message by using a Generative AI Chatbot, or whether Alex has written their message on their own. The text could be entirely generated by a Generative AI Chatbot, or it could be completely written by Alex, word by word.*"

Each participant was randomly assigned to one of the four above conditions, thus our experiment had a full-factorial 4x4 design, with two factors (type of scenario and type of information about AI use) that we manipulated between participants, resulting in 16 unique scenario-information combinations.

Next, we asked participants to report their first impression of Alex as an open-ended response. Participants could write as much or as little as they preferred. We then asked participants to report their impressions of Alex along 10 dimensions (see Section 2.4). Following the social impression evaluations, participants answered a series of questions about their understanding of, and familiarity with Generative AI; their own use of Generative AI chatbots; and their perceptions about the prevalence of Generative AI chatbot use within the general population and among close others. Finally, we collected participants' demographic information (age, gender, highest level of education, and annual household income). We report the full Qualtrics survey including all the above items and instructions in Supplementary A. Participants who completed the full survey received $1 for their participation, regardless of how they responded to the survey questions.

## 2.4. Measures

Our main dependent measure was participants' social impression of Alex, the sender. We quantified this by asking participants to rate Alex along the following 10 dimensions of personal traits: friendly, sincere, polite, authentic, trustworthy, intelligent, skilled, creative, diligent, and thoughtful. Participants indicated their responses using 7-point Likert scales from *Not at all* (1) to *Very* (7). As pre-registered, we standardized each of these 10 measures within each scenario (to account for potential differences between scenarios) and then created a composite measure of *overall social impression* by averaging the 10 standardized measures. We also assessed the internal consistency of this composite measure using Cronbach's alpha, which yielded a standardized coefficient of α = .97, indicating very high reliability.

After participants evaluated the sender, we asked them a set of five questions measuring their understanding of, and familiarity with, Generative AI tools. Specifically, we asked participants to rate: their familiarity with various tools that utilize Generative AI (from *1 = Not familiar at all* to *7 = Very familiar*); their understanding of Generative AI in general (from *1 = No understanding* to *7 = Expert-level understanding*); their confidence in explaining the concept of Generative AI to others (from *1 = Not confident at all* to *7 = Very confident*); their familiarity with Generative AI chatbots that can generate text (from *1 = Not familiar at all* to *7 = Very familiar*); and their confidence in writing effective prompts when using Generative AI chatbots (from *1 = Not confident at all* to *7 = Very confident*). We also assessed the internal consistency of these five items using Cronbach's alpha. This analysis revealed a standardized coefficient of α = .93, suggesting very high reliability, therefore we created a composite measure of *Generative AI expertise* by averaging the standardized versions of these five measures.

Next, we asked participants to report how often they used Generative AI chatbots in the past two weeks: 1) for any purpose (0-14 days); and 2) specifically for writing or editing messages sent to others (0-14 days).[3]

Finally, to gauge participants' perceptions of the prevalence of *others* using Generative AI chatbots when writing messages, we asked participants to provide an estimate for the following: 1) the proportion of people in the U.S. general population who have *ever* used Generative AI chatbots to generate or edit text sent to others; 2) the proportion of people in the U.S. population who are doing so relatively frequently (at least once per week); 3) the proportion of their close others (i.e., "friends, relatives, close colleagues, close classmates, etc.") who have ever used Generative AI chatbots to generate or edit text sent to others; and 4) the proportion of their close others who are doing so relatively frequently (at least once per week). Participants indicated their responses to all four questions using continuous slider scales from 0% to 100%.

---

[3] If participants responded "0" to the first question, they skipped the second question, and we automatically coded their responses to the second questions as "0" too.

## 3. Results

### 3.1. Actual (self-reported) use of Generative AI chatbots

Before testing our main hypothesis and addressing our primary research questions, we investigated participants' self-reported frequency of using Generative AI chatbots. The majority (72.0%) reported that they have used Generative AI chatbots within the past two weeks ($M$ = 3.76 days; median = 2 days).[4] About a fifth of the total sample (22.1%) have used such tools at least every other day (7 or more days), with 4% using these tools every single day (see Figure 1). This indicates that as of November 2024, the use of Generative AI chatbots is already quite widespread among the general population in the United States, consistent with the findings of Bick et al. (2024).

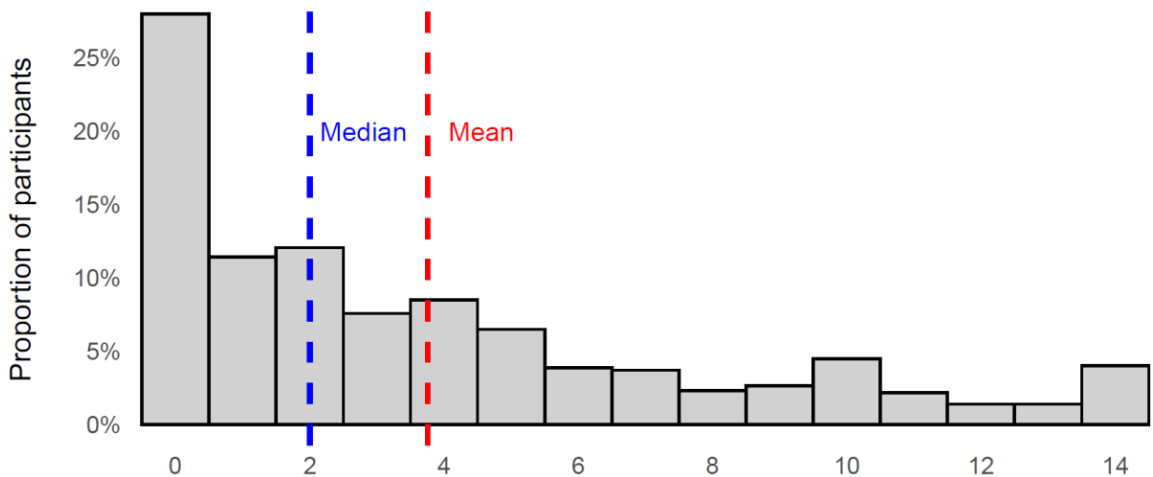

*Figure 1.* Participants' self-reported use of Generative AI chatbots (for any purpose) within the past two weeks (number of days), $N$ = 647.

---

[4] Even though we collected these self-reported measures after participants were exposed to the main experimental manipulation (i.e., whether the hypothetical message was human-written or AI-generated), we did not find any significant effects of condition on these measures, all $p \geq .292$. This suggests that this experimental manipulation is unlikely to affect participants' self-reports about their own use of Generative AI chatbots.

However, these reports include *all purposes* of using Generative AI, including those that do not involve any subsequent communication with another person (e.g., asking questions, generating content for own entertainment, testing chatbots, etc.). To get a better idea about how often people use these tools for communication purposes, we also examined participants' self-reported use of Generative AI chatbots specifically for *generating or editing messages that they sent to someone else*. Slightly less than half of the sample (45.7%) reported that they have used Generative AI within the past two weeks to send a message to someone else ($M = 2.25$ days; median = 0 days). Among these users, 37.9% (of the total sample) utilized the technology at least twice during this period (approximately weekly), 13.6% used it at least every other day (7 or more days), and 1.2% sent an AI-generated or AI-edited message on a daily basis (see Figure 2).

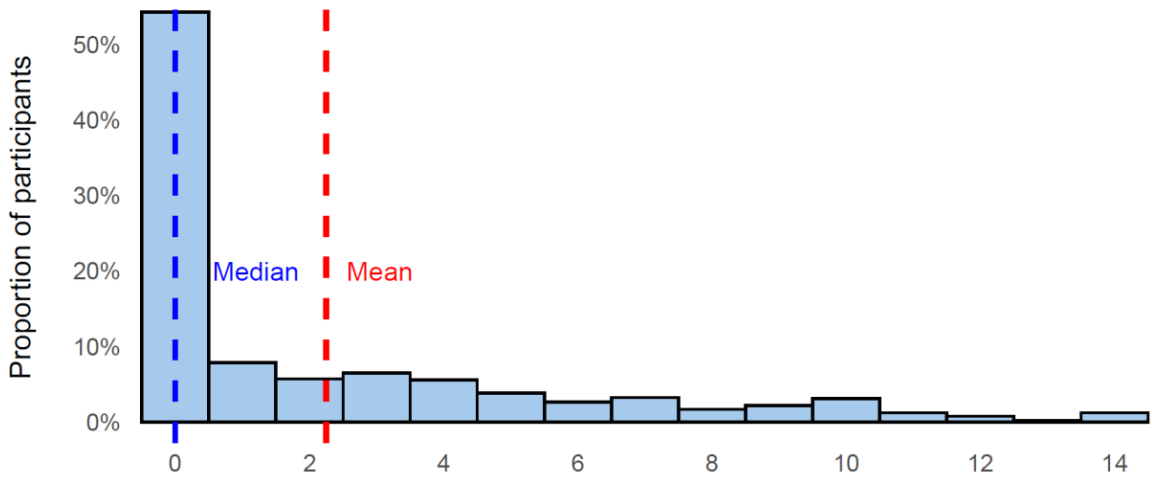

*Figure 2.* Participants' self-reported use of Generative AI chatbots (for writing messages to others) within the past two weeks (number of days), $N = 647$.

While these proportions are substantially lower than the proportions we observed for generic uses of Generative AI chatbots, it is noteworthy that a non-negligible proportion of the study sample (over 10%) reported using such tools to generate or edit messages sent to others on a regular basis (*at least* every other day), and close to half of the sample have sent such messages at least once within the past two weeks.

*3.2. Who is using Generative AI chatbots for writing messages?*

To better understand participants' use of Generative AI chatbots—especially when used for writing messages to others—we investigated whether the self-reported use was associated with the demographic factors we collected (i.e., age, gender, income, education, self-reported expertise/familiarity with Generative AI tools).[5] We found that self-reported use of Generative AI chatbots (for any purpose) was significantly positively correlated with participants' self-reported expertise/familiarity with these tools, $r(646) = .527$, $p < .001$, and significantly negatively correlated with participants' age, $r(646) = -.107$, $p = .006$. We also observed a significant positive relationship between chatbot use and education level, $r(645) = .087$, $p = .027$. Participants' gender and income did not significantly correlate with their self-reported use of Generative AI chatbots, both $p \geq .280$.

Similarly, we found that self-reported use of Generative AI chatbots for generating messages was significantly positively correlated with participants expertise/familiarity with these tools, $r(646) = .362$, $p < .001$, and significantly negatively correlated with participants' age, $r(646) = -.105$, $p = .007$. We also observed a significant positive relationship between chatbot use for writing messages and education level, $r(645) = .184$, $p < .001$. We did not find a significant correlation between self-reported use for writing text and gender, $p = .192$, however, interestingly, we found a significant positive correlation with income, $r(631) = .110$, $p = .006$. This suggests that while higher income participants were not more likely to use Generative AI chatbots *in general*, they were significantly more likely to report using such tools for *writing messages* to others. For example, participants who reported household incomes over $100,000 within the past year used these tools to write messages 48.5% more often ($M = 2.78$ days), compared to participants who reported incomes below $50,000 ($M = 1.87$ days), even though their overall use of Generative AI chatbots was only 7.8% higher ($M = 3.81$ days vs. $M = 3.54$ days; see Figure 3).

---

[5] We report the results of Pearson correlation analyses, except for the correlations between education level and other variables. Since education is an ordinal variable, we report the results of Spearman rank-ordered correlation analyses whenever we discuss associations between highest level of education and other measures. We report the full results of all Pearson and Spearman correlation analyses in Supplementary B.

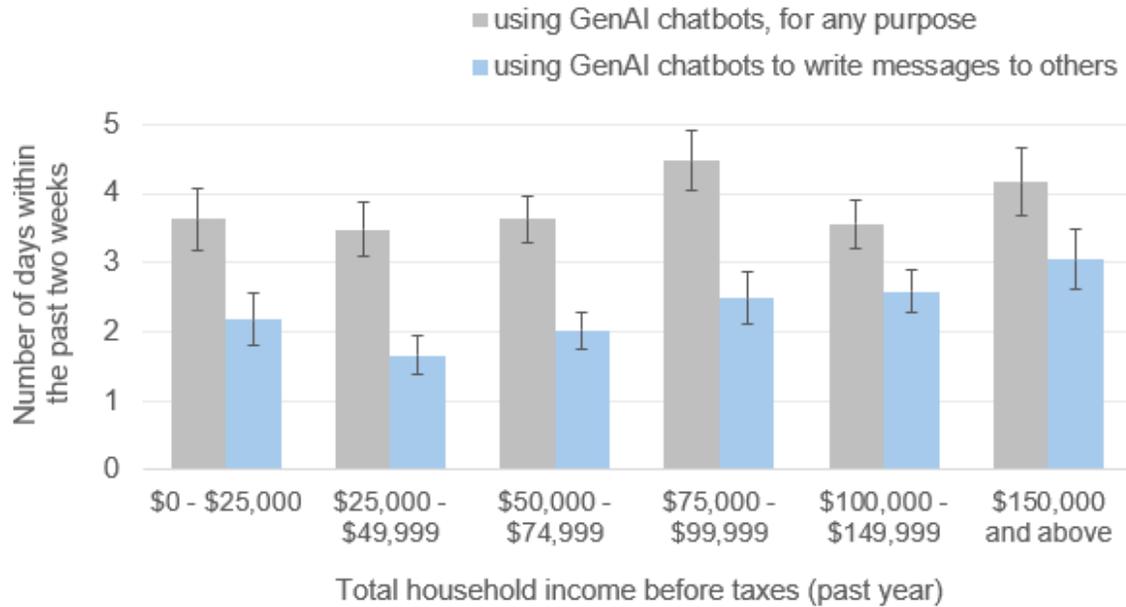

*Figure 3.* Participants' self-reported use of Generative AI chatbots within the past two weeks (number of days), based on their reported household income, $N = 632$ (15 participants preferred not to disclose their income; we omitted these participants from this analysis). Error bars represent ±1 standard error.

We also found a similar pattern when comparing the use of Generative AI across different levels of education. While there were relatively modest differences in the overall use of Generative AI chatbots, participants with more advanced degrees reported significantly more frequent use of these tools for writing messages to others. For example, participants with advanced degrees (professional or doctoral degrees) reported using Generative AI chatbots for any purpose at a similar frequency ($M = 3.54$ days) as those without a college degree ($M = 3.30$ days). However, they reported using such tools for writing messages nearly twice as often ($M = 2.96$ days vs. $M = 1.51$ days; see Figure 4). To summarize, we found that younger, more educated, and higher income participants were particularly likely to use Generative AI chatbots to write messages to others, even though these participants were not necessarily more likely to use such tools in general.

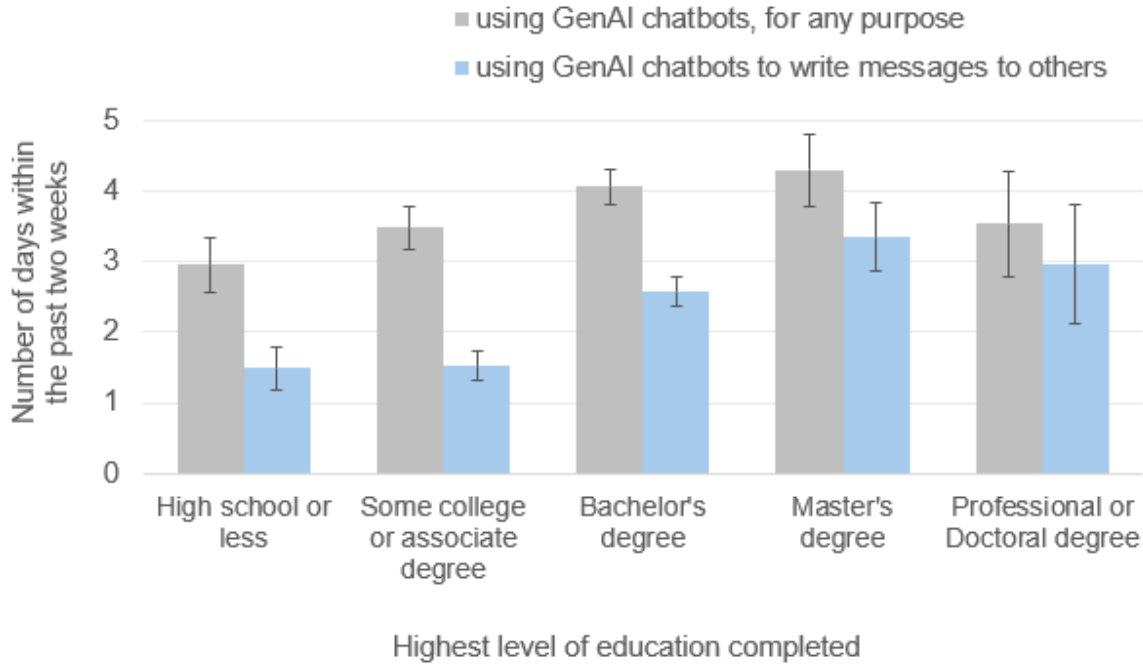

*Figure 4.* Participants' self-reported use of Generative AI chatbots within the past two weeks (number of days), based on their reported education, $N = 646$ (one participant preferred not to disclose their education level; we omitted this participant from this analysis). Error bars represent ±1 standard error.

## 3.3. Perceptions of others' use of Generative AI chatbots for writing messages

While we found that a substantial proportion of the sample reported using Generative AI chatbots for writing messages to others, our results suggest that people consistently underestimate how widespread the use of these tools is when it comes to writing and editing messages. Participants on average believed that only 42.4% ($SD = 22.1\%$) of the U.S. general population has *ever* used such tools for writing or editing messages sent to others, which is significantly lower than the proportion of the present sample who self-reported doing so *just within the past two weeks* (45.7%), $t(646) = 3.86$, $p < .001$, $d = 0.15$, 95% CI [40.69, 44.10]. People also underestimated the proportion of the population who are using these tools relatively frequently (at least on a weekly basis) to write or edit their messages: On average, participants thought that only 31.6% ($SD = 24.4\%$) of the U.S. population is doing so, which is significantly less than the proportion we observed in the current sample (37.9%), $t(646) = 6.49$, $p < .001$, $d = 0.26$, 95% CI [29.76, 33.53].

Interestingly, when asked about their perception of their close others' (friends, relatives, close colleagues, close classmates, etc.) use of Generative AI chatbots, participants thought that their close others are substantially less likely to use these tools for writing and editing messages than the general U.S. population. On average, participants thought that only 34.0% ($SD = 28.7\%$) of their close others have ever done so, and only 24.8% ($SD = 27.0\%$) have done so on a regular (weekly) basis, both of which are significantly lower than the corresponding perceptions about the general population (42.4% and 31.6%, respectively), both $p < .001$. This suggests that people may be particularly naive when judging how widespread the adoption of these tools is within their close social network.

Participants' self-reported use of Generative AI chatbots significantly positively correlated with their perceptions about others using such tools, all $r \geq .200$, all $p < .001$, while we found a significant negative correlation between participants' age and their perceptions, all $r \leq -.094$, all $p \leq .017$. We found mixed and weak associations between participants' perceptions of others' use of Generative AI chatbots and participants' gender, income, and education level, which we report in detail in Supplementary B.

## 3.4. Main results: Social impressions

We conducted independent samples (two-tailed) $t$-tests to compare the overall social impressions across conditions.[6] Confirming our hypothesis (***H1***), participants in the human condition evaluated the sender significantly more positively ($M = 0.39$) than participants in the AI condition ($M = -0.94$), $t(245) = 15.28$, $p < .001$, $d = 1.69$, 95% CI [1.16, 1.51], see Figure 5. Participants also evaluated the sender significantly more positively in the no information condition ($M = 0.43$) and the uncertain condition ($M = 0.13$) compared to the AI condition (***RQ2A*** and ***RQ2B***, respectively), $t(234) = 15.93$, $p < .001$, $d = 1.76$, 95% CI [1.20, 1.54] and $t(284) = 11.52$, $p < .001$, $d = 1.28$, 95% CI [0.89, 1.26], respectively. Overall impressions were also significantly more positive in the human condition, compared to the uncertain condition (***RQ3B***), $t(303) = 3.91$, $p < .001$, $d = 0.43$, 95% CI [0.13, 0.39].

---

[6] As pre-registered, we applied a Bonferroni correction when conducting these $t$-tests to adjust for multiple comparisons, except for the test between the Human and AI conditions, since we made a directional, a priori hypothesis about the latter effect. The $p$ values reported in this section are the Bonferroni-adjusted values.

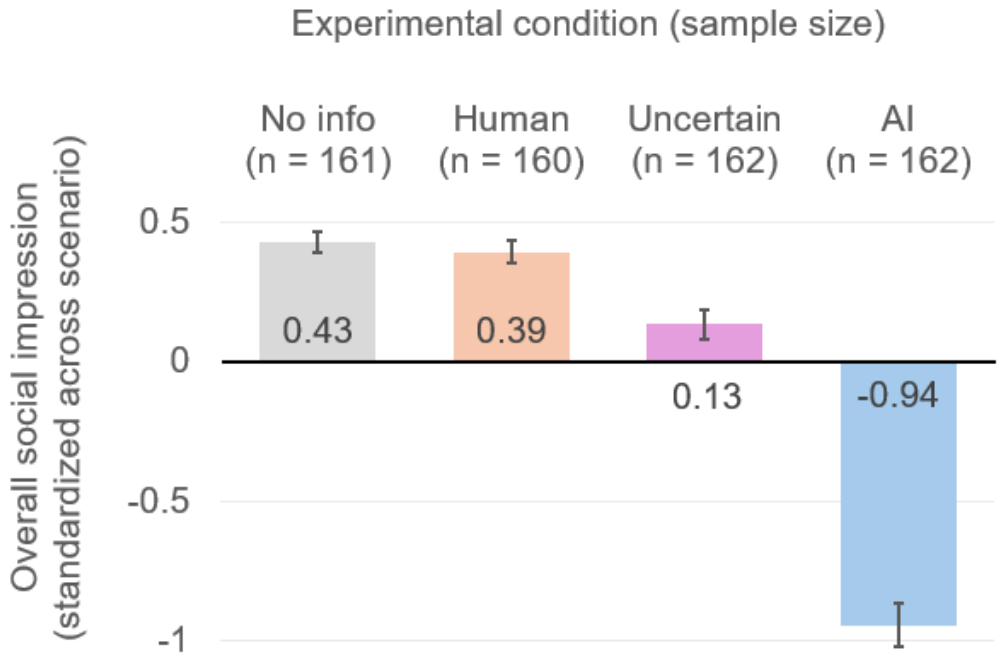

*Figure 5.* Overall social impression across conditions. The overall social impression is the average of the 10 items of social impression (e.g., friendly, intelligent, polite). We standardized the 10 individual items within each scenario. We also report these results separately by scenario in Supplementary C. Error bars represent ±1 standard error.

By contrast, we did not observe a significant difference between the human condition and the no information condition (***RQ3A***), $t(318) = 0.61$, $p = 1$, $d = 0.07$, 95% CI [-0.14, 0.08]. This suggests that when people lack any explicit information about the sender's potential use of Generative AI, they assume that the message has been entirely written by the sender. This result is especially striking given that almost half of the sample admitted having used such tools for writing and editing messages sent to others, just within the past two weeks.

Interestingly, we found a significant difference between the no information and uncertain conditions (***RQ4***), even though we didn't provide any definite information about the sender's use of Generative AI in the latter condition. Participants evaluated the sender significantly more positively in the no information condition than in the uncertain condition, $t(291) = 4.54$, $p < .001$, $d = 0.51$, 95% CI [0.17, 0.42]. This suggests that when the possibility of Generative AI use is highlighted—even if it cannot be determined with certainty—people become more critical and judge the writer more harshly.

It is important to note that even in the uncertain condition, overall impressions are much closer to the human condition than to the AI condition (cf. effect sizes in the corresponding pairwise comparisons: $d = 0.43$ vs. $d = 1.28$), which suggests that participants judged the message to be more likely human-written than AI-generated. If we assume that the overall impression in the uncertain condition is the linear combination of the impression judgments in the AI and human conditions, weighted by corresponding probabilities (see Equation 1), then we can calculate the average perceived probability of Generative AI use by rearranging Equation 1:

$$p_{AI} = (I_U - I_H)/(I_{AI} - I_H)$$

*Equation 2.* Implied probability that the message was generated by AI in the uncertain condition, based on the average impressions in the uncertain, human, and AI conditions.

Solving the above formula, we get $p_{AI} = 19.4\%$, which suggests that people thought the message was about four times more likely to be human-written (79.6%) than AI-generated `(19.4%). We obtain similar results in all four scenarios for $p_{AI}$: 18.4%, 23.8%, 14.3%, and 20.2%, in the gratitude, nanny, cover letter, and feedback scenarios, respectively.

### 3.5. *Exploratory sentiment analysis of open-ended first impressions*

Participants in the human condition wrote overwhelmingly positive descriptions of the sender (e.g., "I think Alex seems very professional, well-prepared, and eloquent" or "Alex seems to be a kind, appreciative, and thoughtful person. Their message is heartful, showing genuine gratitude and warmth"). By contrast, in the AI conditions, these first impressions took a noticeably more negative tone (e.g., "I would find Alex to be lazy and insincere" or "I would be very upset and hurt that Alex didn't even take the time to thank me himself after everything I did for him, and was so lazy he used AI").

To quantify these differences in first impressions and to compare the average sentiments across conditions, we performed lexicon-based sentiment analysis at the word level using the Syuzhet package in R (Jockers, 2017). In this package, sentiment is determined by counting the total number of negative and positive words in a corpus of text, identified by the lexicon (i.e., the sentiment score of a given text is the number of positive words minus the number of negative words). After we obtained the sentiment scores for each response, we compared the average sentiments (i.e., net valence) across conditions, using independent samples (two-tailed) *t*-tests.

These tests revealed that the sentiment of first impressions was significantly more positive in the human condition ($M = 0.110$) than in the AI condition ($M = 0.029$), $t(7390) = 11.76$, $p < .001$, $d = 0.26$, 95% CI [0.067, 0.094]. Participants' first impression in the human condition contained both significantly more positive words ($M = 0.118$) and significantly fewer negative words ($M = 0.008$) than in the AI conditions ($M = 0.061$ and $M = 0.032$, for positive and negative words, respectively), both $p < .001$ (see Figure 6). Overall sentiment was also significantly more positive in the no information ($M = 0.114$) and uncertain ($M = 0.099$) conditions than in the AI condition, both $p < .001$. However, we didn't observe any significant differences between the no information, human, and uncertain conditions, all $p \geq .206$.

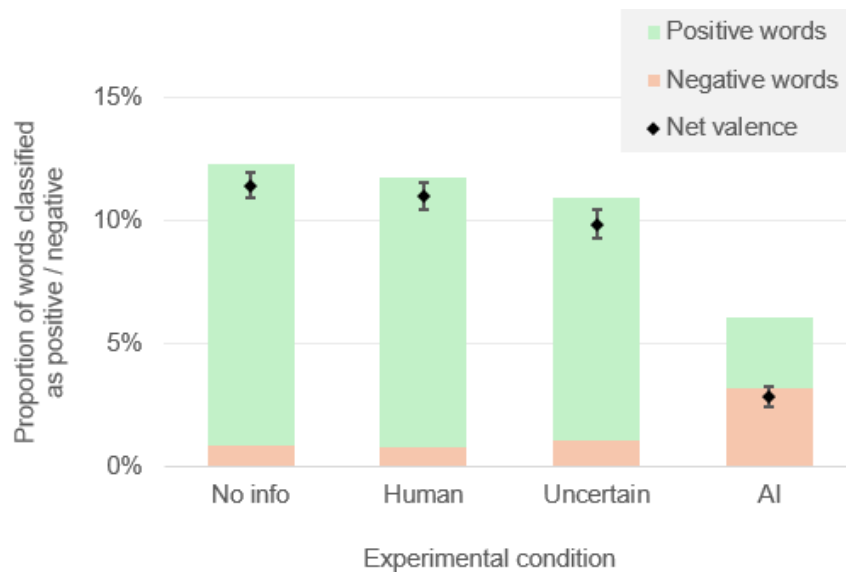

*Figure 6.* Results of the sentiment analysis of the open-ended first impressions across conditions. Values indicate the proportion of words classified as positive (green) or negative (orange). The black diamond markers indicate the average net sentiments. Error bars represent ±1 standard error.

We also calculated the implied probability that the message was generated by AI in the uncertain condition, using the formula described in Equation 2 (but replaced impressions with sentiments). This analysis revealed an implied probability of $p_{AI} = 13.6\%$, which is even lower than the value we obtained using participants' explicit social impression ratings (19.4%). This suggests that participants' first impressions—as reflected by the overall sentiment of the words they used—were even more optimistic in the uncertain condition than their explicit ratings, and indistinguishable from their first impressions in the no information and human conditions.

## 4. Discussion

Consistent with earlier work (Glikson & Asscher, 2023; Hohenstein et al., 2023; Lim et al., 2025; Weiss et al., 2022), we found strong negative effects on social impressions of the sender when disclosing that a message was generated by AI, compared to when the same message was believed to be written by a human. We detected these effects across four distinct communication scenarios (personal vs. professional; close others vs. strangers) and in both participants' explicit numeric ratings of the sender (i.e., overall social impression scale) and their implicit sentiments towards the sender (i.e., net valence of their open-ended impressions).

However, when the involvement of Generative AI was less than certain, participants formed overly positive impressions of the sender. In particular, when we did not even highlight the possibility that the message was generated by AI, participants' impressions of the sender were virtually indistinguishable from the impressions they formed when they *knew* that the message was fully written by a human. Interestingly, we found significant differences in the explicit ratings between the uninformed and uncertain conditions, which suggests that participants, by default, did not even consider the possibility of AI involvement, but adjusted their impressions once this possibility was highlighted. Crucially, we did not tell participants any definite information about the origin of the message in either of these conditions, so the observed differences can only be explained by different levels of attention to AI involvement. Similarly to earlier work on the link between social judgments of behavior and attention to others' vaccination status during the COVID-19 pandemic (Molnar et al., 2023), here we document that social judgments can be significantly affected by shifting people's attention to decision-relevant information that otherwise does not readily come to mind.

At the same time, even when we prompted participants to consider the possibility that a message was fully generated by AI, the resulting social impressions were much closer to those of human-written messages, as opposed to AI-generated messages. In addition, participants seem to underestimate the prevalence of Generative AI chatbot use, especially among their close peers, even though almost half of the current study sample admitted using such tools for writing messages to others, *just within the past two weeks*. In the following sections, we discuss some potential alternative explanations for these results. Exploring which of these mechanisms is responsible for the main effects observed in our experiment is beyond the scope of the current paper but could be a suitable topic for future research.

## 4.1. Lack of knowledge and experience with Generative AI

The first possibility is that people may be unaware of the capabilities of these emerging technologies and simply would not think that Generative AI can produce messages of this quality (even though we generated these messages using ChatGPT-4o and only minimally edited them for length and consistency across scenarios). Note, however, that we did not find any significant correlations between participants' self-reported AI expertise and familiarity with these tools and their social impressions (see Supplementary B), which suggests that the lack of knowledge may not be the main culprit here. In addition, even if people are fully aware of what these technologies are capable of, they often cannot distinguish between human-written and AI-generated content (Köbis & Mossink, 2021; Kreps et al., 2022) and may rate the AI-generated content as "more human than human" (Jakesch et al., 2023; Porter & Machery, 2024). Finally, due to the very recent emergence of these tools (ChatGPT was initially released in November 2022: less than two years before we conducted this experiment), most people may simply lack the first-hand experience of receiving an authentic-looking message that later turns out to be AI-generated. As the rich literature on the description-experience gap in risky decision-making (e.g., Haines et al., 2023; Hertwig & Erev, 2009; Wulff et al., 2018) highlights, people may react to indirect, abstract information (such as reading about a scenario in an online study, as in our experiment) very differently to direct information that they obtain through personal experience (e.g., if people would suspect one of their actual colleagues to have generated an email). For example, Chen et al. (2018) demonstrated that participants became better calibrated in trusting a phishing detection system only when they got direct experience along with feedback, rather than merely reading descriptions of such a system. Similarly, it is possible that people would become better calibrated (i.e., form more negative impressions of others when they don't know whether others used Generative AI) after being personally exposed to such situations in real life.

## 4.2. Asymmetric costs of social errors

When we calculated the implied probability in the uncertain condition that the message was written by Generative AI (cf. Equation 2), we relied on the assumption that participants would follow the logic of *expected value*: that the impression formed under uncertainty is the linear additive combination of impressions made under certainty (AI and human conditions), weighted by the corresponding probabilities. However, this may not be a valid assumption.

Another possibility is that people deviate from the expected value framework and prefer to err on the safe side (trusting the sender when they are uncertain) if the social costs of errors are highly asymmetric. According to error management theory (see Haselton & Buss, 2000; McKay & Efferson, 2010), outcomes of social interactions may carry asymmetric costs, and when people face such situations, they may prefer to *minimize such costs*, rather than trying to maximize the expected value of the outcome. Translating this to the present context, there are two types of errors one can make: believing that the message was genuine when it was actually generated by AI or believing that the message is AI-generated when it was actually written by the sender. It is possible that people perceive the latter type of error (i.e., a false accusation) socially more costly than the former type of error (i.e., naivete), thus they may adopt an "innocent until proven guilty" attitude when forming impressions of others based on their writing.

## 4.3. Motivated cognition

Finally, it is possible that people are *motivated* to believe that others are less likely to use Generative AI chatbots when writing messages, compared to the actual prevalence of this behavior, and this may be reflected in their overly positive impressions in the uninformed and uncertain conditions. People are often motivated to form and maintain overly optimistic beliefs that make them feel good about themselves, other people, and the world in general (see, e.g., Bénabou & Tirole, 2016; Epley & Gilovich, 2016; Molnar & Loewenstein, 2021). When these motives are strong enough, people are even willing to sacrifice the accuracy of their judgments, in order to increase the valence of their beliefs (Golman et al., 2017, 2022; Sharot & Sunstein, 2020; Sweeny et al., 2010). People may experience stronger AI anxiety (Johnson & Verdicchio 2017; Li & Huang, 2020) if they believe that the use of Generative AI chatbots is widespread, or they may feel disappointed, angry, and disconnected when they suspect others of using such tools for writing messages, especially their close others. To mitigate these negative emotions, people may adopt overly optimistic beliefs about others' use of Generative AI chatbots. In the present study, we do find suggestive evidence for such a mechanism when comparing participants' belief about the use of Generative AI chatbots among their close social peers, as opposed to the general U.S. population: Participants expected substantially lower prevalence among people they knew well compared to the population at large. This is consistent with the motive to think more positively about socially relevant close others, as opposed to strangers.

## 5. Conclusion

We found that despite the relatively widespread use of Generative AI in writing or editing messages, participants did not exhibit *any* skepticism towards senders under naturalistic evaluation conditions, when the possibility of AI use was not explicitly highlighted. Under these realistic conditions, participants' impressions of senders were virtually indistinguishable from impressions of senders who were *known* to have written their messages on their own, without using Generative AI. Even when we highlighted the potential use of Generative AI, participants formed overly positive impressions of senders, especially in their open-ended first impressions. Notably, when participants were certain that the sender used Generative AI, their impressions were substantially more critical than in any other condition, so the overly positive judgments in the uninformed and uncertain conditions cannot be explained by participants' overall indifference towards the use of Generative AI. Instead, our results suggest that while judgments can be substantially affected by information about Generative AI use, this information is not readily available and not top of mind in more realistic communication settings.

These findings put existing work in a new light by challenging the prevailing assumptions about the negative impact of AI on social perceptions. Previous studies have documented adverse effects of AI involvement in communication, such as reduced authenticity and sincerity in apologies and other messages. However, our research suggests that when the source of a message is uncertain, individuals may overlook the potential downsides of AI and default to more favorable evaluations. This indicates a nuanced understanding of how uncertainty and attention to AI involvement can shape social judgments, highlighting the need for further exploration into the conditions under which skepticism may arise.

Looking ahead, it is essential to consider how these dynamics may evolve over time. As awareness of Generative AI's capabilities and its implications for communication increases, individuals may begin to adopt a more skeptical perspective. Furthermore, long-term effects of widespread use of Generative AI could lead to a shift in social perceptions, prompting individuals to critically evaluate the authenticity and credibility of messages they receive. New technological advances (e.g., more efficient detection, watermarking, etc.) and regulatory policies may also influence public perception and trust toward written communication. Future research should explore these potential changes in attitudes and beliefs, particularly as society becomes more accustomed to the integration of AI technologies in everyday communication.

## Credit author statement

**Jiaqi Zhu:** Conceptualization, Methodology, Software, Formal analysis, Investigation, Resources, Writing - Original Draft, Writing - Review & Editing, Visualization

**Andras Molnar:** Conceptualization, Methodology, Software, Validation, Formal analysis, Investigation, Data Curation, Writing - Original Draft, Writing - Review & Editing, Visualization, Supervision, Project administration, Funding acquisition

## Declaration of generative AI and AI-assisted technologies in the writing process

During the preparation of this work the authors used ChapGPT-4o in order to generate the initial drafts of the messages used in the four hypothetical scenarios. After using this tool, the authors reviewed and revised these messages as needed. The authors did not use generative AI or AI-assisted technologies during any other part of the writing process, and take full responsibility for the content of the published article.

## Funding sources

This research did not receive any specific grant from funding agencies in the public, commercial, or not-for-profit sectors.

## Data availability statement

All data, analysis code, and research materials are publicly available at OSF: https://osf.io/xrpy9/?view_only=ba3a2e85211140fdb005ed21de92b2a7.

# References

Appelbaum, M., Cooper, H., Kline, R. B., Mayo-Wilson, E., Nezu, A. M., & Rao, S. M. (2018). Journal article reporting standards for quantitative research in psychology: The APA Publications and Communications Board task force report. *American Psychologist*, *73*(1), 3–25. https://doi.org/10.1037/amp0000191

Bénabou, R., & Tirole, J. (2016). Mindful Economics: The Production, Consumption, and Value of Beliefs. *Journal of Economic Perspectives, 30*(3), 141–164. https://doi.org/10.1257/jep.30.3.141

Bick, A., Blandin, A., & Deming, D. (2024). The Rapid Adoption of Generative AI (*NATIONAL BUREAU OF ECONOMIC RESEARCH No. w32966*). Cambridge, MA. https://doi.org/10.3386/w32966

Blease, C. R., Locher, C., Gaab, J., Hägglund, M., & Mandl, K. D. (2024). Generative artificial intelligence in primary care: an online survey of UK general practitioners. *BMJ Health & Care Informatics, 31*(1), e101102. https://doi.org/10.1136/bmjhci-2024-101102

Chaudhry, S. J., & Wald, K. A. (2022). Overcoming listener skepticism: Costly signaling in communication increases perceived honesty. *Current Opinion in Psychology, 48*, 101442. https://doi.org/10.1016/j.copsyc.2022.101442

Chen, J., Mishler, S., Hu, B., Li, N., & Proctor, R. W. (2018). The description-experience gap in the effect of warning reliability on user trust and performance in a phishing-detection context. *International Journal of Human-Computer Studies, 119*, 35–47. https://doi.org/10.1016/j.ijhcs.2018.05.010

Connelly, B. L., Certo, S. T., Ireland, R. D., & Reutzel, C. R. (2011). Signaling Theory: A Review and Assessment. *Journal of Management, 37*(1), 39–67. https://doi.org/10.1177/0149206310388419

Dedema, M., & Ma, R. (2024). The collective use and perceptions of generative AI tools in digital humanities research: Survey-based results. https://doi.org/10.48550/arXiv.2404.12458

Enke, B. (2020). What You See Is All There Is. *The Quarterly Journal of Economics, 135*(3), 1363–1398. https://doi.org/10.1093/qje/qjaa012

Epley, N., & Gilovich, T. (2016). The Mechanics of Motivated Reasoning. *Journal of Economic Perspectives, 30*(3), 133–140. https://doi.org/10.1257/jep.30.3.133

Faul, F., Erdfelder, E., Buchner, A., & Lang, A.-G. (2009). Statistical power analyses using G*Power 3.1: Tests for correlation and regression analyses. *Behavior Research Methods, 41*(4), 1149–1160. https://doi.org/10.3758/BRM.41.4.1149

Gao, C., Chen, D., Zhang, Q., Huang, Y., Wan, Y., & Sun, L. (2024). LLM-as-a-Coauthor: The Challenges of Detecting LLM-Human Mixcase. *arXiv preprint arXiv:2401.05952*. https://doi.org/https://doi.org/10.48550/arXiv.2401.05952

Gintis, H., Smith, E. A., & Bowles, S. (2001). Costly Signaling and Cooperation. Journal of Theoretical Biology, 213(1), 103–119. https://doi.org/10.1006/jtbi.2001.2406

Glikson, E., & Asscher, O. (2023). AI-mediated apology in a multilingual work context: Implications for perceived authenticity and willingness to forgive. *Computers in Human Behavior, 140*, 107592. https://doi.org/10.1016/j.chb.2022.107592

Golman, R., Hagmann, D., & Loewenstein, G. (2017). Information Avoidance. *Journal of Economic Literature, 55*(1), 96–135. https://doi.org/10.1257/jel.20151245

Golman, R., Loewenstein, G., Molnar, A., & Saccardo, S. (2022). The Demand for, and Avoidance of, Information. *Management Science, 68*(9), 6454–6476. https://doi.org/10.1287/mnsc.2021.4244

Haines, N., Kvam, P. D., & Turner, B. M. (2023). Explaining the description-experience gap in risky decision-making: learning and memory retention during experience as causal mechanisms. *Cognitive, Affective, & Behavioral Neuroscience, 23*(3), 557–577. https://doi.org/10.3758/s13415-023-01099-z

Haselton, M. G., & Buss, D. M. (2000). Error management theory: A new perspective on biases in cross-sex mind reading. *Journal of Personality and Social Psychology, 78*(1), 81–91. https://doi.org/10.1037/0022-3514.78.1.81

Heintz, C., Karabegovic, M., & Molnar, A. (2016). The Co-evolution of Honesty and Strategic Vigilance. *Frontiers in Psychology, 7*, 1–13. https://doi.org/10.3389/fpsyg.2016.01503

Hertwig, R., & Erev, I. (2009). The description–experience gap in risky choice. *Trends in Cognitive Sciences, 13*(12), 517–523. https://doi.org/10.1016/j.tics.2009.09.004

Hohenstein, J., Kizilcec, R. F., DiFranzo, D., Aghajari, Z., Mieczkowski, H., Levy, K., … Jung, M. F. (2023). Artificial intelligence in communication impacts language and social relationships. *Scientific Reports, 13*(1), 5487. https://doi.org/10.1038/s41598-023-30938-9

Humlum, A., & Vestergaard, E. (2024). The Adoption of ChatGPT. SSRN Electronic Journal, (202). https://doi.org/10.2139/ssrn.4827166

Jakesch, M., Hancock, J. T., & Naaman, M. (2023). Human heuristics for AI-generated language are flawed. *Proceedings of the National Academy of Sciences, 120*(11), 2017. https://doi.org/10.1073/pnas.2208839120

Jockers, M. (2017). Package 'syuzhet'. https://cran.r-project.org/web/packages/syuzhet

Johnson, D. G., & Verdicchio, M. (2017). AI Anxiety. *Journal of the Association for Information Science and Technology, 68*(9), 2267–2270. https://doi.org/10.1002/asi.23867

Kahneman, D. (2011). Thinking, fast and slow. Farrar, Straus and Giroux.

Köbis, N., & Mossink, L. D. (2021). Artificial intelligence versus Maya Angelou: Experimental evidence that people cannot differentiate AI-generated from human-written poetry. *Computers in Human Behavior, 114*, 106553. https://doi.org/10.1016/j.chb.2020.106553

Kreps, S., McCain, R. M., & Brundage, M. (2022). All the News That's Fit to Fabricate: AI-Generated Text as a Tool of Media Misinformation. *Journal of Experimental Political Science, 9*(1), 104–117. https://doi.org/10.1017/XPS.2020.37

Li, J., & Huang, J.-S. (2020). Dimensions of artificial intelligence anxiety based on the integrated fear acquisition theory. *Technology in Society*, 63, 101410. https://doi.org/10.1016/j.techsoc.2020.101410

Lim, J. S., Schneider, E., Grover, M., Zhang, J., & Peters, D. (2025). Effects of AI versus human source attribution on trust and forgiveness in the identical corporate apology statement for a data breach scandal. *Public Relations Review, 51*(1), 102520. https://doi.org/10.1016/j.pubrev.2024.102520

Mairesse, F., Walker, M. A., Mehl, M. R., & Moore, R. K. (2007). Using Linguistic Cues for the Automatic Recognition of Personality in Conversation and Text. *Journal of Artificial Intelligence Research, 30*, 457–500. https://doi.org/10.1613/jair.2349

McKay, R., & Efferson, C. (2010). The subtleties of error management. *Evolution and Human Behavior, 31*(5), 309–319. https://doi.org/10.1016/j.evolhumbehav.2010.04.005

Molnar, A., & Loewenstein, G. F. (2021). Thoughts and Players: An Introduction to Old and New Economic Perspectives on Beliefs. *In J. Musolino, J. Sommer, & P. Hemmer (Eds.), The Cognitive Science of Beliefs.* Cambridge University Press. https://doi.org/10.2139/ssrn.3806135

Molnar, A., Moore, A., Fowler, C., & Wu, G. (2023). Seen and not seen: How people judge ambiguous behavior during the COVID-19 pandemic. *Journal of Risk and Uncertainty, 66*(2), 141–159. https://doi.org/10.1007/s11166-022-09396-7

Pennebaker, J. W., Mehl, M. R., & Niederhoffer, K. G. (2003). Psychological Aspects of Natural Language Use: Our Words, Our Selves. *Annual Review of Psychology, 54*(1), 547–577. https://doi.org/10.1146/annurev.psych.54.101601.145041

Porter, B., & Machery, E. (2024). AI-generated poetry is indistinguishable from human-written poetry and is rated more favorably. *Scientific Reports, 14*(1), 26133. https://doi.org/10.1038/s41598-024-76900-1

R Core Team (2024). R: A Language and Environment for Statistical Computing. R Foundation for Statistical Computing, Vienna, Austria. https://www.R-project.org

Sadasivan, V. S., Kumar, A., Balasubramanian, S., Wang, W., & Feizi, S. (2023). Can AI-generated text be reliably detected? *arXiv preprint arXiv:2303.11156.* https://doi.org/10.48550/arXiv.2303.11156

Sharot, T., & Sunstein, C. R. (2020). How people decide what they want to know. *Nature Human Behaviour, 4*(1), 14–19. https://doi.org/10.1038/s41562-019-0793-1

Simon, H. A. (1955). A Behavioral Model of Rational Choice. *The Quarterly Journal of Economics, 69*(1), 99. https://doi.org/10.2307/1884852

Slovic, P. (1972). From Shakespeare to Simon: Speculations and some evidence. *Oregon Research Institute Bulletin, 12*(2), 1–19.

Spence, M. (1973). Job Market Signaling. *The Quarterly Journal of Economics, 87*(3), 355. https://doi.org/10.2307/1882010

Spence, M. (2002). Signaling in Retrospect and the Informational Structure of Markets. *American Economic Review, 92*(3), 434–459. https://doi.org/10.1257/00028280260136200

Sweeny, K., Melnyk, D., Miller, W., & Shepperd, J. A. (2010). Information avoidance: Who, what, when, and why. *Review of General Psychology, 14*(4), 340–353. https://doi.org/10.1037/a0021288

Tang, R., Chuang, Y.-N., & Hu, X. (2024). The Science of Detecting LLM-Generated Text. *Communications of the ACM, 67*(4), 50–59. https://doi.org/10.1145/3624725

Tomlinson, B., Black, R. W., Patterson, D. J., & Torrance, A. W. (2024). The carbon emissions of writing and illustrating are lower for AI than for humans. *Scientific Reports, 14*(1), 3732. https://doi.org/10.1038/s41598-024-54271-x

Van Noorden, R., & Perkel, J. M. (2023). AI and science: what 1,600 researchers think. *Nature, 621*(7980), 672–675. https://doi.org/10.1038/d41586-023-02980-0

Weiss, D., Liu, S. X., Mieczkowski, H., & Hancock, J. T. (2022). Effects of Using Artificial Intelligence on Interpersonal Perceptions of Job Applicants. *Cyberpsychology, Behavior, and Social Networking, 25*(3), 163–168. https://doi.org/10.1089/cyber.2020.0863

Wells, G. L., & Windschitl, P. D. (1999). Stimulus Sampling and Social Psychological Experimentation. *Personality and Social Psychology Bulletin, 25*(9), 1115–1125. https://doi.org/10.1177/01461672992512005

Wojtowicz, Z., & DeDeo, S. (2025). Undermining Mental Proof: How AI Can Make Cooperation Harder by Making Thinking Easier. *arXiv preprint arXiv:2407.14452*. http://arxiv.org/abs/2407.14452

Wulff, D. U., Mergenthaler-Canseco, M., & Hertwig, R. (2018). A meta-analytic review of two modes of learning and the description-experience gap. *Psychological Bulletin, 144*(2), 140–176. https://doi.org/10.1037/bul0000115

Zhang, J. S., Yoon, C., Williams, D. K. A., & Pinkas, A. (2024). Exploring the Usage of ChatGPT Among Medical Students in the United States. *Journal of Medical Education and Curricular Development, 11*, 1–7. https://doi.org/10.1177/23821205241264695

Zhu, T., Zhang, K., & Wang, W. Y. (2024). Embracing AI in Education: Understanding the Surge in Large Language Model Use by Secondary Students, 1–8. Retrieved from http://arxiv.org/abs/2411.18708

## Appendix: Hypothetical scenarios used in the experiment

### *"Gratitude" scenario*

[Context]

"Your close friend Alex broke their leg a month ago while playing sports. You drove Alex to the hospital immediately and have been taking care of them ever since. You visited Alex as many times as you could. Almost every day, you've been bringing food or cooking for Alex. You also helped with daily tasks for Alex to make their life easier, and handled some of the work and personal responsibilities they are unable to manage during recovery.

Today, you received the following email from Alex:"

[Message (230 words)]

"My Dearest Friend,

I've been reflecting on all the incredible ways you've supported me over the past month, and I simply must take a moment to express my heartfelt gratitude. When I broke my leg, I felt so lost and helpless, but you stepped in like a ray of sunshine, always there to lift me up. From racing me to the hospital with such urgency to bringing me delicious meals every day, you've been my rock. Your kindness in helping with my work and managing the daily tasks I couldn't handle has truly touched my heart.

Thank you for being by my side during this challenging time; your company turned my recovery into a journey filled with warmth and laughter. I can't begin to describe how much easier and brighter you've made this period for me.

I honestly don't know how I would have navigated this without you. Your generosity and thoughtfulness mean the world to me. It's in moments like these that I'm reminded of how incredibly fortunate I am to have you in my life.

Please know that your support hasn't gone unnoticed. I am forever grateful for everything you've done. If you ever find yourself in need, I promise I'll be there for you, just as you've been there for me. I cherish our friendship more than words can say.

With all my love and gratitude,
Alex"

### *"Nanny" scenario*

[Context]

"You are looking to hire a nanny for your family, so you posted an advertisement to attract potential applicants. Several individuals have been responding to your ad via email, typically providing a written self-introduction along with a description of their experience. You plan to make your hiring decision based on these initial email interactions.

Today, you received the following email from one of the candidates, Alex:"

[Message (217 words)]

"Dear family,

I hope this message finds you well! My name is Alex, and I'm reaching out with great enthusiasm about the nanny position you've posted. With years of experience working with children, I would be thrilled to become a part of your family's daily routine.

Throughout my journey as a nanny, I've had the joy of caring for children of all ages, from tiny babies to energetic school-age kids. Each experience has been incredibly rewarding, and I cherish the moments spent helping children learn and grow through creative activities, outdoor adventures, and nurturing play. Whether it's whipping up healthy meals, assisting with homework, or simply being a comforting presence, I take great pride in supporting both the kids and parents I work with.

I'm also more than happy to lend a hand with light housework or running errands when needed. My ultimate goal is to create a warm, structured environment where your children can thrive, all while making life a little easier for your family.

I would love the opportunity to chat and learn more about your family and what you're looking for in a nanny. Thank you so much for considering my application—I'm genuinely excited about the possibility of working for your family and contributing to a positive and nurturing environment!

Warmest regards, Alex"

### *"Cover letter" scenario*

[Context]

"You work in the Human Resources (HR) department of a big media company that has recently opened a position for a Data Analyst. Your team has been receiving applications from various candidates. In their cover letters, job-seekers usually introduce themselves, provide an overview of their relevant experience and express their enthusiasm for the role they are applying for.

Today, you received the following cover letter from one of the candidates, Alex B.:"

[Message (239 words)]

"Dear Madam or Sir,

I am writing to formally express my interest in the Data Analyst position. With over two years of experience in data analysis, I have developed a robust skill set in data visualization, statistical modeling, and delivering actionable insights that align with business strategies.

In my previous roles, I successfully managed a variety of projects that involved analyzing large datasets, constructing predictive models, and collaborating with cross-functional teams. My work has focused on uncovering trends and providing data-driven recommendations to inform key decision-making processes. Notably, I led a project that involved building predictive models aimed at enhancing customer retention, resulting in a marked increase in retention rates in the next season. Additionally, I regularly engaged with other departments to present findings and optimize operational workflows through data insights.

I am particularly impressed by your company's commitment to innovation and the critical role that data plays in shaping business decisions. I am confident that my analytical expertise and passion for utilizing data to address complex challenges would make me a valuable asset to your team. Furthermore, I have long admired your company's strong reputation, dynamic culture, and collaborative environment, and I am eager to contribute to its continued success.

Thank you for considering my application. I would appreciate the opportunity to discuss how my experience and skills can align with the needs of your position. Please feel free to contact me at your convenience.

Sincerely,
Alex B."

## *"Feedback" scenario*

[Context]

"You are working at a marketing firm as a member of the team led by the senior manager of the department. You have recently completed the first draft of a project proposal for a consumer satisfaction study. Alex is a junior manager in a different team, but has some relevant expertise in consumer behavior and marketing strategy. You want to make sure your proposal is polished and effective before submitting it, so you reached out to Alex for some feedback.

Today, you received the following email from Alex:"

[Message (243 words)]

"Hi,

I've had the opportunity to thoroughly review your proposal, and I want to commend you on the solid foundation you've established. It's evident that you've invested considerable thought and effort into crafting this document, which is a great start. That said, I believe there are several key areas where we could enhance the proposal:

1. Focus: I recommend refining the focus to spotlight a select few critical drivers that have a direct correlation with consumer loyalty and retention. This targeted approach could yield deeper insights and strengthen your overall argument.
2. Segmentation: I would encourage you to consider integrating behavioral segmentation as well. This addition could significantly enrich the insights provided, enabling us to develop more nuanced strategies that resonate with distinct consumer groups.
3. Survey design: I believe incorporating a few open-ended questions could be beneficial. This would allow for the capture of qualitative insights that may reveal unexpected pain points or opportunities that predefined questions might overlook.
4. Strategic implications: I encourage you to frame your findings in a manner that directly ties back to actionable business strategies. It would be advantageous to illustrate how the insights garnered can lead to concrete improvements in consumer satisfaction.

If you have any questions or would like to delve deeper into any of these suggestions, please feel free to reach out. I would be more than happy to collaborate further and assist in refining the proposal to its fullest potential.

Best regards,
Alex"

The following information was displayed to all participants in the human, AI, and uncertain conditions, immediately after they read the message:

Now, we would like you to learn about **Generative AI Chatbots**.

A Generative AI Chatbot is a **computer program that can understand and generate human-like text**. It uses advanced machine learning techniques to have conversations, answer questions, and provide information on a wide range of topics. People may use Generative AI Chatbots to **generate all kinds of written messages in all sorts of contexts**, and these texts can be indistinguishable from human-written text.

Learn more about Generative AI chatbots (link opens in a new window)

In the no information condition, participants read the above text <u>after</u> they reported their impressions, during the post-decision survey (i.e., when reporting their own attitudes and use of Generative AI).